\documentclass[12pt,titlepage]{article}
\usepackage[dvips]{graphicx}
\begin{document}

\title{\textbf{The Blackbody Radiation\\
in $D$-Dimensional Universes}\thanks{To appear in Revista Brasileira de Ensino de F\'{\i}sica}\\
{\small (A Radia\c{c}\~{a}o de Corpo Negro em Universos
$D$-Dimensionais)}}
\author{Tatiana R. Cardoso and Antonio S. de Castro\thanks{
Enviar correspond\^{e}ncia para A. S. de Castro. E-mail: castro@feg.unesp.br.%
} \\
\\
UNESP - Campus de Guaratinguet\'{a}\\
Departamento de F\'{\i}sica e Qu\'{\i}mica\\
Caixa Postal 205\\
12516-410 Guaratinguet\'{a} SP - Brasil}
\date{}
\date{}
\maketitle

\begin{abstract}
\noindent The blackbody radiation is analyzed in universes with $D$ spatial
dimensions. With the classical electrodynamics suited to the universe in
focus and recurring to the hyperspherical coordinates, it is shown that the
spectral energy density as well as the total energy density are sensible to
the dimensionality of the universe. Wien's displacement law and the
Stefan-Boltzmann law are properly generalized. \newline
\vspace{.05cm} \newline
\vspace{1cm} \newline
\centerline{\textbf{Resumo}}\vspace{0.3cm} \noindent A radia\c{c}\~{a}o de
corpo negro \'{e} analisada em universos com $D$ dimens\~{o}es espaciais.
Usando a eletrodin\^{a}mica cl\'{a}ssica consonante com a dimensionalidade
do universo, e recorrendo ao sistema de coordenadas hiperesf\'{e}ricas,
mostra-se que tanto a densidade de energia espectral quanto a densidade de
energia total s\~{a}o sens\'{\i}veis ao n\'{u}mero de dimens\~{o}es
espaciais. A lei do deslocamento de Wien e a lei de Stefan-Boltzmann s\~{a}o
apropriadamente ge\-ne\-ralizadas. \newline
\vspace{.05cm} \newline
\noindent \textbf{Palavras-chave:} corpo negro, $D$ dimens\~{o}es, dimens%
\~{o}es extras, coordenadas hiperesf\'{e}ricas \newline
\end{abstract}

\section{Introduction}

The dimensionality of space can be specified by the number $D$, where $D+1$
is the maximum number of points which can be mutually equidistant. The
dimensionality of the space we live, the physical space, is not derived from
any physical law but we perceive that we can move ourselves in three
distinct directions: to the left and to the right, forward and backward,
upward and downward. Therefore, it is convenient for most of the purposes to
take for granted that we live in a 3-dimensional universe. Obviously this
perception of dimensionality is conditioned by the limitations of our
sensations, and a possible world with more than three dimensions might be
properly detected by means of experiments realized in laboratories.

A number of theories involving the unification of the fundamental physical
interactions, such as Kaluza-Klein theories \cite{coq} and superstring
theories \cite{gre}-\cite{riv}, demand physical universes encompassing extra
spatial dimensions. It is argued that those extra dimensions do not manifest
themselves neither to our sensory experiences nor to the laboratory
experiments because such extra dimensions are curled up into dimensions of
the order of $10^{-33}$cm, an extremely small size to be aware of through
the senses or even to be detected by experiments. Physical theories in lower
dimensions are also interesting not only as simpler models for the
``re\-a\-listic'' three-dimensional world but also for exhibiting some
peculiarities that turn out them different from the 3-dimensional physical
theories \cite{abd}-\cite{eft}.

The growing interest in the physics of multidimensional universes calls for
a pedagogical approach of a few topics of physics that are usually explored
only in the three-dimensional space. The classical electrodynamics in
universes with 1 and 2 spatial dimensions has already received a didactic
approach by Lapidus \cite{lap}. The author also speculated about the
classical electrodynamics in $D$-dimensional universes. Influenced by Lapidus%
\'{}s work \cite{lap}, the present paper approaches the influence of the
dimensionality of space in the blackbody radiation, one of the first topics
the students face at their beginning studies on quantum theory.

The present analysis of the blackbody radiation in $D$-dimensional universes
follows the standard procedure for counting the number of standing wave
modes in a cavity (see, \textit{e.g.}, \cite{fey}-\cite{eis}). Nevertheless,
the classical electrodynamics consonant with the dimensionality of the
universe in question is used. As a part of the discourse, a few criticisms
to the literature is also presented. Recurring to the hyperspherical
coordinates, it is shown that both the spectral energy density and the total
energy density are sensible to the dimensionality of the universe. Wien\'{}s
displacement law as well as the Stefan-Boltzmann law are properly
generalized. Finally, the readers are instigated to generalize the
calculation of the specific heat capacity of a crystalline solid and the
calculation of the average energy of the particles of an ideal Bose gas in $%
D $-dimensional universes.

\section{Preliminary precautions}

Lapidus \cite{lap} stated that the electric field in a one-dimensional
universe obeys the wave equation. That statement induces unwary readers to
guess about the propagation of electric waves in such a universe. To tell
the truth, neither there are electric waves nor electromagnetic waves. In
such an one-dimensional universe the magnetic field is absent and ``Maxwell%
\'{}s equations'' are given by \cite{lap}
\begin{equation}
\frac{\partial E}{\partial x}=\frac{\rho }{\varepsilon _{0}},\qquad \frac{%
\partial E}{\partial t}=-\,\frac{J}{\varepsilon _{0}}  \label{1}
\end{equation}

\noindent where $\varepsilon _{0}$ is the permittivity of free space. The
charge and current densities satisfy the equation of continuity: $\partial
\rho /\partial t+\partial J/\partial x=0$. In the absence of sources the
only possible configuration for the electric field, $E$, is to be
homogeneous and static. A homogeneous and static electric field always
satisfies the wave equation in a trivial way but it is clear there is no
electromagnetic wave in an one-dimensional universe. The absence of magnetic
fields in such an one-dimensional universe reinforces our argument. This
initial observation naturally restricts the considerations of the cavity
radiation for universes with $D>1$.

It is true that Refs. \cite{fey}-\cite{eis} take into account the
electromagnetic modes in a cubic cavity, though, as a didactic appeal, they
first consider the modes in a one-dimensional cavity. It is worthwhile to
mention that such an one-dimensional cavity does not consist in an
one-dimensional world, but only an one-dimensional cavity embedded in a
three-dimensional world. This argument is strengthen by noting that the
authors consider two possible directions for the polarization of the
electric field. This observation refuses the statement of the authors of
Ref. \cite{eis}, when they remark that the number of modes in
one-dimensional and three-dimensional cavities depend on different powers
for the frequencies just because the dimensionality of the corresponding
universes differ.

The observations in the previous paragraphs illustrate the subtle and
crucial role of the dimensionality of the universe concerning the
propagation of electromagnetic waves.

\section{The cavity radiation in D-dimensional universes}

The law of the blackbody radiation is customarily formulated as the
electromagnetic energy density inside a cavity into the frequency range
between $\nu $ and $\nu +d\nu $. Thus, one must know the density of states
in the interval $d\nu $ and multiply it by the average energy.

Let us consider the electromagnetic radiation in thermal equilibrium in a
metallic cavity with the shape of a $D$-dimensional cube (hypercube\footnote{%
Square in the case $D=2$, cube in the case $D=3$.}) whose walls are
maintained at the absolute temperature $T$. Choosing a system of orthogonal
Cartesian coordinates with origin at one of the vertex of the cube and axes
parallel to their edges, one can write the components of the electric field
corresponding to the standing waves in the cavity as
\begin{equation}
E_{a}(x_{a},t)=E_{a}^{(0)}\sin (|\kappa _{a}|x_{a})\sin (2\pi \nu t)
\label{2}
\end{equation}

\noindent where $E_{a}^{(0)}$ is the amplitude of the $a$-th component of
the electric field oscillating with frequency $\nu $, $\kappa _{a}$ is the
component of the\ wave number, and $a=1,2,\ldots D$. The electric field must
vanish at the walls of the cavity so that the components of the wave number
satisfy the conditions
\begin{equation}
|\kappa _{a}|=\frac{\pi }{L}\,n_{a},\quad n_{a}=1,2,3,\ldots  \label{3}
\end{equation}

\noindent where $L$ is the length of the edge of the cube. Therefore, the
number of possible modes for the electric field in a given state of
polarization, into the interval between $\kappa _{a}$ and $\kappa
_{a}+d\kappa _{a}$, is given by
\begin{equation}
N_{P}(\kappa _{a})d\kappa _{a}=\frac{L}{2\pi }\,d\kappa _{a}  \label{4}
\end{equation}

\noindent The presence of the factor 2 in the denominator of (\ref{4})
allows us to consider that the components of the vector wave number have
positive as well as negative values.

The total number of modes in the cavity is the product of the possible modes
in each Cartesian axis:
\begin{equation}
N_{P}(|\vec{\kappa}|)d|\vec{\kappa}|=\prod_{a=1}^{D}N_{P}(\kappa
_{a})d\kappa _{a}=V\,\frac{dV_{\kappa }}{\left( 2\pi \right) ^{D}}\,
\label{5}
\end{equation}

\noindent where $V=L^{D}$ is the volume of the cavity and $dV_{\kappa }$ is
the infinitesimal $D$-dimensional element of volume in the $\kappa $-space,
\textit{viz.}
\begin{equation}
dV_{\kappa }=\prod_{a=1}^{D}d\kappa _{a}  \label{5-1}
\end{equation}
This element of volume is, as a matter of fact, a volume which fill the
space between the ($D-1$)-dimensional hyperspherical shells\footnote{%
Circles in the case $D=2$, spherical surfaces in the case $D=3$.} centered
about the origin with radii $|\vec{\kappa}|$ and $|\vec{\kappa}|+d|\vec{%
\kappa}|$.

In a $D$-dimensional world the electric field has $D$ components and the
magnetic field has $D\left( D-1\right) /2$ components \cite{lap}. Since the
electromagnetic waves are\ transverse, one can conclude that the electric
field has $D-1$ mutually perpendicular components standing at right angles
to the direction of the wave propagation. In other words, there are $D-1$
possible directions of polarizations for the electric field\footnote{%
If the precautions foreseen in Section 2 were not considered, one might
conclude that there are no electromagnetic waves in an one-dimensional
universe but there might be longitudinal electric waves. This last
possibility, though, only can be excluded by the analysis of Maxwell\'{}s
equations expressed by (1).}. In this way, the number of possible modes,
including all possible polarizations of the electric field, is
\begin{equation}
N(|\vec{\kappa}|)d|\vec{\kappa}|=\left( D-1\right) \,V\,\frac{dV_{\kappa }}{%
\left( 2\pi \right) ^{D}}  \label{6}
\end{equation}

\noindent To evaluate $dV_{\kappa }$ it is convenient to recur to the
hyperspherical coordinates. The hyperspherical coordinates in a space with $%
D $ dimensions ($D>1$) are related to the Cartesian coordinates by \cite{erd}%
:

\begin{eqnarray}
x_{1} &=&r\cos \theta _{1}\sin \theta _{2}\ldots \sin \theta _{D-1}
\nonumber \\
&&  \nonumber \\
x_{2} &=&r\sin \theta _{1}\sin \theta _{2}\ldots \sin \theta _{D-1}
\nonumber \\
&&  \label{7} \\
x_{a} &=&r\cos \theta _{a-1}\sin \theta _{a}\ldots \sin \theta _{D-1},\quad
\mathrm{for}\quad 3\leq a\leq D-1  \nonumber \\
&&  \nonumber \\
x_{D} &=&r\cos \theta _{D-1}  \nonumber
\end{eqnarray}

\noindent in such a way that the distance of a point to the origin is
expressed in terms of the Cartesian coordinates as $r=\sqrt{%
\sum_{a=1}^{a=D}x_{a}^{2}}$. In addition,
\begin{equation}
0\leq \theta _{1}\leq 2\pi ,\quad \mathrm{and}\quad 0\leq \theta _{a}\leq
\pi \quad \mathrm{for}\quad 2\leq a\leq D-1  \label{8}
\end{equation}
\noindent The $D$-dimensional element of volume\footnote{%
Element of area in the case $D=2$.} and the ($D-1$)-dimensional element of
solid angle\footnote{%
Element of plane angle in the case $D=2$.} are given by

\begin{equation}
\prod_{a=1}^{D}dx_{a}=r^{D-1}dr\,d\Omega ,\qquad d\Omega
=\prod_{a=1}^{D-1}\left( \sin \theta _{a}\right) ^{a-1}d\theta _{a}
\label{9}
\end{equation}

\noindent For calculating the total solid angle, instead of proceeding with
the calculation of the integrals by brute force, some special functions are
used. The beta function, $B(z,w)$, can be defined as \cite{abr}
\begin{eqnarray}
B(z,w) &=&2\int_{0}^{\pi /2}\left( \sin \theta \right) ^{2z-1}\left( \cos
\theta \right) ^{2w-1}\,\,d\theta  \nonumber \\
&&  \label{9-1} \\
&=&\frac{\Gamma (z)\Gamma (w)}{\Gamma (z+w)},\quad \mathrm{Re}(z)>0,\quad
\mathrm{Re}(w)>0  \nonumber
\end{eqnarray}
where $\Gamma (z)$ is the gamma function
\begin{equation}
\Gamma (z)=\int_{0}^{\infty }y^{z-1}e^{-y}\,\,dy  \label{10-1}
\end{equation}
Then, knowing that $\Gamma (1/2)=\sqrt{\pi }$, one can write
\begin{equation}
\int_{0}^{\pi }\left( \sin \theta _{a}\right) ^{a-1}\,\,d\theta _{a}=\sqrt{%
\pi }\,\frac{\Gamma \left( \frac{a}{2}\right) }{\Gamma \left( \frac{a+1}{2}%
\right) }  \label{10-2}
\end{equation}
in such a manner that the total solid angle is given by
\begin{equation}
\Omega =2\pi ^{D/2}\times \left\{
\begin{array}{c}
1 \\
\\
\prod_{a=2}^{a=D-1}\frac{\Gamma \left( \frac{a}{2}\right) }{\Gamma \left(
\frac{a+1}{2}\right) }%
\end{array}
\begin{array}{l}
,\quad \mathrm{for}\quad D=2 \\
\\
,\quad \mathrm{for}\quad D\geq 3%
\end{array}
\right.  \label{10-3}
\end{equation}
Writing $\Omega $ in the compact form
\begin{equation}
\Omega =\frac{2\,\pi ^{D/2}}{\Gamma (D/2)}  \label{10}
\end{equation}

\noindent

\noindent one can finally express $dV_{\kappa }$ as
\begin{equation}
dV_{\kappa }=\frac{2\pi ^{D/2}}{\Gamma \left( D/2\right) }\,|\vec{\kappa}%
|^{D-1}d|\vec{\kappa}|  \label{11}
\end{equation}

\noindent Note that if the signs of de $\kappa _{a}$ were restricted only to
positive values, the factors $2$ and $2^{D}$ in the denominators of (\ref{4}%
) and (\ref{6}), respectively, would be absent. In this circumstance only
the infinitesimal element of volume in the space $\kappa $ corresponding to $%
\kappa _{a}>0$ should be considered, implying that the result expressed by (%
\ref{11}) should be divided by $2^{D}$ for restrict itself to a $2^{D}$-ant%
\footnote{%
A quadrant in the case $D=2$, an octant in the case $D=3$, and so on.}.

Since $|\vec{\kappa}|=2\pi \nu /c$, where $c$ is the speed of the
electromagnetic wave, the number of modes with frequencies between $\nu $
and $\nu +d\nu $ can be written as
\begin{equation}
N(\nu )d\nu =\left( D-1\right) \,V\,\frac{2}{\Gamma \left( D/2\right) }%
\,\left( \frac{\sqrt{\pi }}{c}\right) ^{D}\,\nu ^{D-1}d\nu  \label{12}
\end{equation}

\noindent Now, each mode has an average energy given by Plank\'{}s radiation
formula independently of the dimensionality of the universe \cite{fey}-\cite%
{eis}, \textit{i.e.},
\begin{equation}
E_{med}=\frac{h\nu }{\exp \left( h\nu /k_{B}T\right) -1}  \label{13}
\end{equation}

\noindent where $h$ and $k_{B}$ are the Planck and Boltzmann constants,
respectively, in such a way that the energy density\footnote{%
Energy per unit area in the case $D=2$, energy per unit volume in the case $%
D=3$, and so on.} into the interval between $\nu $ and $\nu +d\nu $ is given
by
\begin{equation}
\rho _{T}\left( \nu \right) d\nu =2\,\,\left( \frac{\sqrt{\pi }}{c}\right)
^{D}\frac{D-1}{\Gamma \left( D/2\right) }\,\frac{h\nu ^{D}}{\exp \left( h\nu
/k_{B}T\right) -1}\,\,d\nu  \label{14}
\end{equation}

\noindent This is the mathematical expression for the Plank blackbody
spectrum in a $D$-dimensional universe. \noindent It gives the spectral
distribution for the energy in the cavity. Fig. 1 illustrates the spectral
energy density (energy density per unit frequency) for the particular cases $%
D=2,3$ and $4$, for $T=1500\,$K\footnote{%
Due to errors on the scales of the figures 1-1 and 1-8 in the Ref. \cite{eis}%
, both at the abscissa and at the ordinate axes, any attemp to compare with
the Fig. 1 is disabled.}.

The result known as Wien\'{}s displacement law, concerned to the
proportionality between the temperature and the frequency that maximizes the
spectrum, \textit{i.e.} $\nu _{\max }/T=\alpha =\mathrm{const}$, is licit
whatsoever the dimensionality of the universe. Nevertheless, the constant $%
\alpha $ increases as the number of the dimensions of the universe increase.
Indeed, substituting $x=h\nu _{\max }/k_{B}T$ in (\ref{14}) and making $%
d\rho _{T}\left( x\right) /dx=0$, one finds
\begin{equation}
e^{-x}=1-\frac{x}{D}  \label{15-1}
\end{equation}

\noindent Eq. (\ref{15-1}) has as positive solution\footnote{$x=0$ is
another possible solution.}
\begin{equation}
x=D+W\left( -De^{-D}\right)  \label{15-2}
\end{equation}

\noindent where $W$ is the Lambert W function defined as \cite{cor}
\begin{equation}
W(z)+e^{W(z)}=z  \label{15-3}
\end{equation}

\noindent Note that if $x$ is to be seen as a function of the dimensionality
of the space for $D>1$, it is a monotonic increasing function. This happens
because the second term at the right hand side of (\ref{15-2}) is a
monotonic increasing function for $D>1$ and tends asymptotically to zero as $%
D\rightarrow \infty $. Of course, the very same conclusion concerning the
behaviour of $\alpha $ as a function of $D$ could be obtained in a simpler
manner by plotting Eq. (\ref{15-1}). It should be mentioned, though, that
the numerical solution of (\ref{15-1}) becomes extremely easy by using
computer algebra softwares which have the Lambert W function built in their
libraries.

The total energy density in the cavity is obtained by integrating the
spectral energy density over all the frequencies, $\rho
_{T}=\int_{0}^{\infty }\rho _{T}\left( \nu \right) d\nu $, resulting in
\begin{equation}
\rho _{T}=a_{D}T^{D+1}  \label{16}
\end{equation}

\noindent Using the identities \cite{abr}
\begin{equation}
\Gamma (z+1)=z\,\Gamma (z)\quad \mathrm{and}\quad \Gamma \left( 2z\right) =%
\frac{2^{2z-1/2}}{\sqrt{2\pi }}\,\Gamma \left( z\right) \Gamma \left( z+%
\frac{1}{2}\right)  \label{16-1}
\end{equation}
the constant of proportionality $a_{D}$ can be written as
\begin{equation}
a_{D}=\left( \frac{2}{hc}\right) ^{D}\,\left( \sqrt{\pi }\right)
^{D-1}k_{B}^{D+1}\,D\left( D-1\right) \,\,\Gamma \left( \frac{D+1}{2}\right)
\,\zeta (D+1)  \label{17}
\end{equation}
where $\zeta (z)$ is the Riemann zeta function \cite{abr}:
\begin{equation}
\zeta (z)=\sum_{n=1}^{\infty }n^{-z},\quad \mathrm{Re}(z)>1  \label{18}
\end{equation}
which can also be written in the form
\begin{equation}
\zeta (z)=\frac{1}{\Gamma (z)}\int_{0}^{\infty }\frac{y^{z-1}}{e^{y}-1}\,\,dy
\label{19}
\end{equation}

The rate of radiation of energy per unit area, known as radiancy, is defined
as the total energy emitted per unit time and per unit area of the cavity
surface. The radiancy and the energy density in the cavity are related by
purely geometric factors. It results that such quantities are proportional,
as one will see.

The energy emitted by an infinitesimal element of area $dA$ of the surface
of the cavity in the interval of time $dt$ occupies a hemisphere ($D-1$%
)-dimensional with radius $cdt$ centered about $dA$. The energy contained in
a cylinder\footnote{%
In the case $D=2$ the cylinder is a rectangle, the element of area is an
element of line and the hemisphere is a semicircle.} of length $cdt$, with
inclination $\theta _{D-1}$ with respect to the Cartesian axis $x_{D}$
perpendicular to the element of area, is
\begin{equation}
dU_{T}=\rho _{T}\,cdt\,dA\cos \theta _{D-1}  \label{19-1}
\end{equation}

\noindent Nevertheless, only the fraction $d\Omega /\Omega $ of the energy
contained in the cylinder propagates in the direction specified by the angle
$\theta _{D-1}$. Therefore, the energy emitted in the direction specified by
the angle $\theta _{D-1}$ is
\begin{equation}
R_{T}\left( \theta _{D-1}\right) =\rho _{T}\,c\,\frac{d\Omega }{\Omega }\cos
\theta _{D-1}  \label{19a}
\end{equation}

\noindent In order to obtain the energy emitted all over possible directions
one must integrate the previous expression over all the angles that sweeps
the hemisphere. There results
\begin{equation}
R_{T}=\frac{\rho _{T}\,c}{\Omega }\times \left\{
\begin{array}{c}
2 \\
\\
\pi \\
\\
2\frac{\left( \sqrt{\pi }\right) ^{D-1}}{D-1}\prod_{a=2}^{a=D-2}\frac{\Gamma
\left( \frac{a}{2}\right) }{\Gamma \left( \frac{a+1}{2}\right) }%
\end{array}
\begin{array}{l}
,\quad \mathrm{for}\quad D=2 \\
\\
,\quad \mathrm{for}\quad D=3 \\
\\
,\quad \mathrm{\ for}\quad D\geq 4%
\end{array}
\right.  \label{19-3}
\end{equation}
that can even be written as
\begin{equation}
R_{T}=\frac{\left( \sqrt{\pi }\right) ^{D-1}}{\Gamma \left( \frac{D+1}{2}%
\right) }\,\frac{\rho _{T}\,c}{\Omega }=\frac{\Gamma \left( \frac{D}{2}%
\right) }{\Gamma \left( \frac{D+1}{2}\right) }\,\frac{c}{2\sqrt{\pi }}\,\rho
_{T}\,  \label{19-2}
\end{equation}

Now one are ready to proclaim the law which relates the radiancy with the
temperature, the generalized Stefan-Boltzmann law:
\begin{equation}
R_{T}=\sigma _{D}T^{D+1}  \label{19-10}
\end{equation}
where the factor $\sigma _{D}$ is the generalized Stefan-Boltzmann constant,
given by
\begin{equation}
\sigma _{D}=\left( \frac{2}{c}\right) ^{D-1}\left( \sqrt{\pi }\right)
^{D-2}\,\frac{k_{B}^{D+1}}{h^{D}}\,D\left( D-1\right) \,\,\Gamma \left(
\frac{D}{2}\right) \,\zeta (D+1)\,  \label{20}
\end{equation}
As a function dependent of the number of the dimensions of space, the
generalized Stefan-Boltzmann constant is a monotonic increasing function.

\section{Conclusions}

Thanks to the hyperspherical coordinates, the analytical treatment of the
blackbody radiation in $D$-dimensional universes has been successful. The
case of one-dimensional cavities was excluded from the context by taking
into account the possible independent directions for the polarization of an
electromagnetic wave.

The experimental results are consistent with the blackbody radiation in a
three-dimensional world. Since higher dimensional universes with such extra
dimensions properly compactified were not considered, the possibility of
higher dimensional universes can not be discarded by contrasting the
experimental data with the analysis done in this work.

It is worthwhile to observe that the results presented in this paper would
be unchanged if one had considered the blackbody radiation as a photon gas
obeying the usual distribution of Bose. This is so because one would had to
consider that each photon with energy $h\nu $ can have $D-1$ possible states
of polarization.

The formalism developed in this work can be used to generalize other topics
of the quantum theory in an easy way. For instance, the calculation of the
vibration modes of the atoms in an elastic solid, or the number of states of
a phonon gas, for the generalization of the Debye theory of the specific
heat capacity of a solid. In this scenario, diversely from the case of
electromagnetic waves, longitudinal acoustic waves are feasible in a
one-dimensional universe. It is anticipated that the specific heat capacity
at high temperatures behaves as $C_{V}=DR$ (the generalized Dulong e Petit
law), where $R$ is the universal gas constant, and that at low temperatures
it behaves as $C_{V}\sim T^{D}$. Another interesting topic is that one
related to the Bose condensation. In this last topic the number of quantum
states of the ideal gas with energy into the interval between $E$ and $E+dE$
is proportional to $E^{D/2-1\,}dE$, a result divergent at $E=0$ in the case $%
D=1$ and independent of $E$ in the case $D=2$. The calculations which lead
to these results, as well as the calculation of the average energy of the
particles of an ideal Bose gas (with a promising problem in the calculation
of the number of particles for $D<3$), are left to the readers.

\vspace{0.8in}

\noindent \textbf{Note added in proof}

After the paper was submitted, we become aware of the more technical
and  quasi-homonymous \textquotedblleft Black-body radiation in
extra dimensions\textquotedblright\ by H. Alnes, F. Ravndal and K.
Wehus (arXiv: quant-ph/0506131), uploaded to the web seventeen days
later.

\vspace{0.4in}

\noindent \textbf{Acknowledgments}

This work was supported in part by means of funds provided by CNPq
and FAPESP.

\newpage

\begin{figure}[th]
\begin{center}
\includegraphics[width=9cm, angle=270]{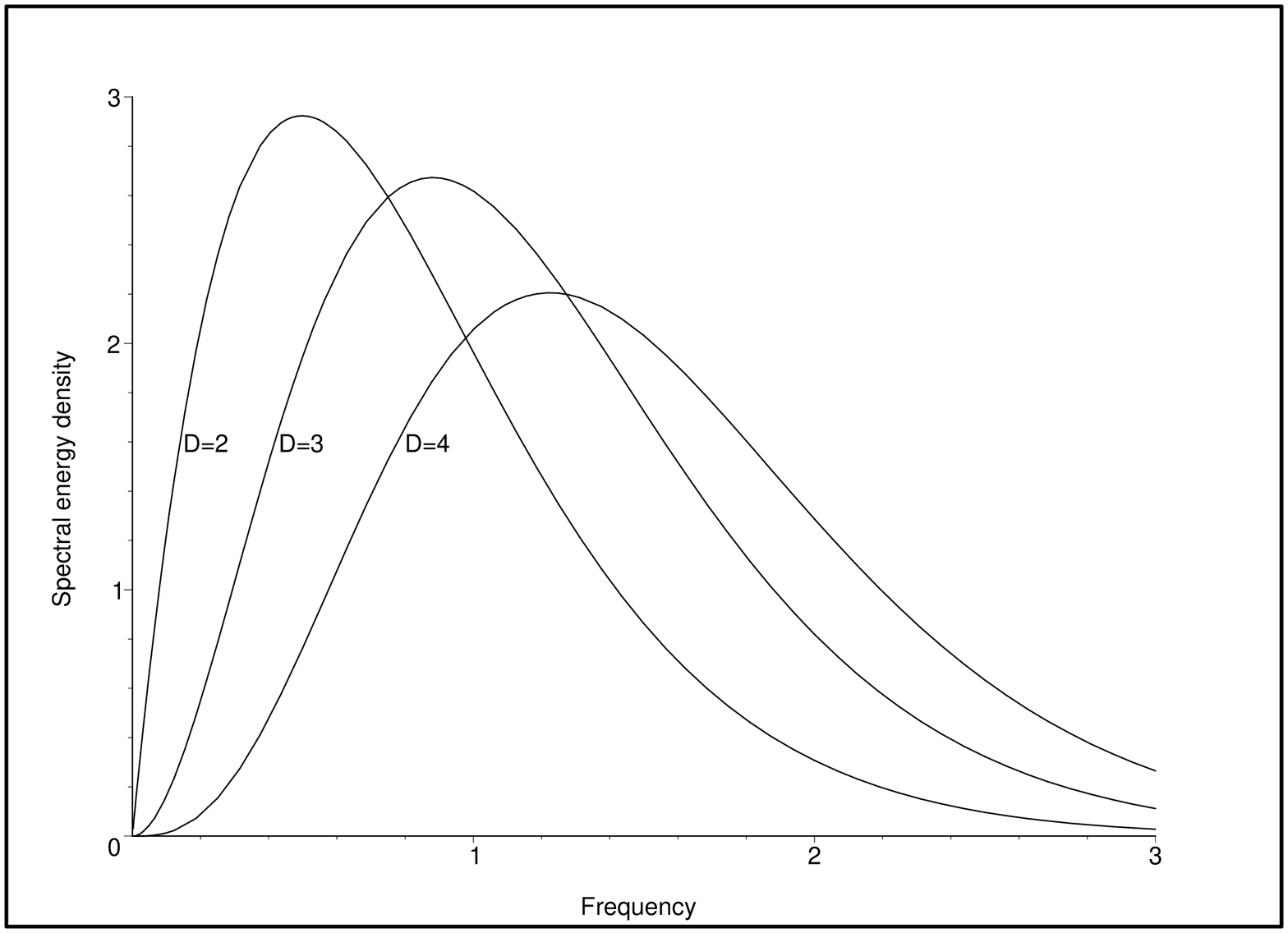}
\end{center}
\par
\vspace*{-0.1cm}
\caption{Spectral energy density as a function of $\protect\nu$ for $T=1500$
K. Using the fundamental physical constants in the international system of
units, the frequency must be multiplied by $10^{14}$ in order to be
expressed in Hz, whereas the spectral energy density must be multiplied by $%
10^{-23}$, $10^{-17}$ and $10^{-11}$ in order to have the units W$\cdot$m$%
^{-1}$$\cdot$Hz$^{-1}$, W$\cdot$m$^{-2}$$\cdot$Hz$^{-1}$ and W$\cdot$m$^{-3}$%
$\cdot$Hz$^{-1}$, in the cases $D=2$, $3$ and $4$, respectively. }
\label{Fig1}
\end{figure}

\end{document}